\documentclass[%
 reprint,
 amsmath,amssymb,
 aps,
prb,
]{revtex4-2}

\usepackage{soul}
\usepackage{enumerate}
\usepackage{amsfonts}
\usepackage{bbm}

\usepackage{graphicx}
\usepackage{dcolumn}
\usepackage{bm}
\usepackage{xcolor}
\usepackage[caption=false]{subfig}
\usepackage[ruled]{algorithm2e}

\newcommand{\dagga}{{\phantom{\dagger}}}

\newenvironment{eqs}%
{\begin{equation} \begin{aligned}}%
{\end{aligned} \end{equation} }
\newcommand{\beal}{\begin{eqs}}
\newcommand{\eal}{\end{eqs}}

 \usepackage{braket}

\begin{document}

\preprint{APS/123-QED}

\title{Quantum embedding for molecules using auxiliary particles – The ghost Gutzwiller Ansatz}
\author{Carlos Mejuto-Zaera}
\email{cmejutoz@sissa.it}
\affiliation{International School for Advanced Studies (SISSA), Via Bonomea 265, 34136 Trieste, Italy}

\date{\today}
\begin{abstract}
Strong/static electronic correlation mediates the emergence of remarkable phases of matter, and underlies the exceptional reactivity properties in transition metal-based catalysts. Modeling strongly correlated molecules and solids calls for multi-reference Ans\"atze, which explicitly capture the competition of energy scales characteristic of such systems. With the efficient computational screening of correlated solids in mind, the ghost Gutzwiller (gGut) Ansatz has been recently developed. This is a variational Ansatz which can be formulated as a self-consistent embedding approach, describing the system within a non-interacting, quasiparticle model, yet providing with accurate spectra in both low and high energy regimes. Crucially, small fragments of the system are identified as responsible for the strong correlation, and are therefore enhanced by adding a set of auxiliary orbitals, the ghosts. These capture many-body correlations through one-body fluctuations and subsequent out-projection when computing physical observables. gGut has been shown to accurately describe multi-orbital lattice models at modest computational cost. In this work, we extend the gGut framework to strongly correlated molecules. To adapt the gGut Ansatz for molecular calculations, we address the fact that, unlike in the lattice model previously considered, electronic interactions in molecules are not local. Hence, we explore a hierarchy of approximations of increasing accuracy capturing interactions between fragments and environment, and within the environment, and discuss how these affect the embedding description of correlations in the whole molecule. We will compare the accuracy of the gGut model with established methods to capture strong correlation within active space formulations, and assess the realistic use of this novel approximation to the theoretical description of correlated molecular clusters.
\end{abstract}
\maketitle

\section{Introduction}
Strong electronic correlation enables the emergence of electronic states with remarkable tuneability~\cite{QuantumMatEditorial}, be it in the form of unconventional superconductors~\cite{Capone2009,Armitage2010,Yin2011,deMedici2014,VillarArribi2021}, ferroelectric perovskites~\cite{Egami1993,Fabrizio1999}, bilayer transition-metal dichalcogenides~\cite{Song2022,Crippa2024} or catalytic molecular complexes~\cite{Chen2008,Weber2013,Weber2014,Lee2022}.
The underlying motif is the competition of multiple energy scales governing the electronic motion, most often potential energy driven localization and kinetic energy favored delocalization~\cite{Imada1998,Lee2006}.
An extensive body of research has been dedicated to capturing strong correlation computationally, resulting in a wide palette of theoretical and numerical approaches aiming to provide the most accurate predictions possible for different observables~\cite{martin_reining_ceperley_2016,Helgaker2000,Schollwock2011,becca2017}.
An important complement to this research lines is the development of simplified models, which while reducing the computational complexity aim to remain qualitatively reliable.
These can be invaluable tools for performing exploratory studies of families of materials, or for proposing phenomenological explanations for correlated behavior.
A successful programme in this direction has been pursued in terms of embedding approximations, such as dynamical mean-field theory (DMFT)~\cite{Kotliar1996,Kotliar2006,Kotliar2001b,Arita2007,Shim2007,Takizawa2009,Haule2010,Park2014,Haule2015,Paul2019}, density-matrix embedding theory (DMET)~\cite{Knizia2012,Wouters2016,Sekaran2021,Sekaran2023}, energy-weighted DMET~\cite{Fertitta2018,Sriluckshmy2021}, or self-energy embedding theory (SEET)~\cite{Kananenka2015,Lan2015,Iskakov2020}, which capture correlation in terms of a small number of orbitals, deemed as the main causes for such correlation.

Embedding approximations vary in their computational complexity and in the range of observables they give access to.
In particular, accessing spectral information, related to ionization potentials and (inverse) photo-emission spectra, is typically associated with a large computational overhead.
Indeed, this usually requires evaluating the one-body Green's function (GF) either directly from a correlated auxiliary impurity model~\cite{Kotliar1996}, or in terms of increasingly complex expectation values~\cite{Fertitta2018}.
Obtaining correlated spectra from an effective, non-interacting model would thus offer a complementary route to evaluating opto-electronic properties of correlated solids and molecules phenomenologically.

Such a route is provided by the ghost Gutzwiller framework (gGut)~\cite{Lanata2017,guerci2019,frank2021,Mejuto2023a,Lee2023a,Lee2023b,Guerci2023}, a generalization of the Gutzwiller Ansatz~\cite{Gutzwiller1963,Gutzwiller1965,Bunemann1998,Fabrizio2007,lanata2015,fabrizio2017} which models correlation in terms of Slater determinants where the weight of expensive charge fluctuations is supressed by local linear operators.
This is also equivalent to an effectively non-interacting quasi-particle Hamiltonian self-consistently coupled to an interacting impurity model.
Crucially, however, upon self-consistency spectral information is evaluated in terms of the quasi-particle Hamiltonian, not the impurity model, making its description computationally inexpensive.
This is achieved by adding auxiliary quasi-particle states to the Ansatz, which are ultimately projected out when computing observables of the physical system.
Further, this embedding framework derives from a variational wave function Ansatz, simplifying the access to forces and electron-vibration couplings as well.

So far, the gGut approach has been successfully applied to investigate lattice models relevant to a family of strongly correlated materials, such as cuprate superconductors, iron pnictides and perovskite materials~\cite{Lanata2017,guerci2019,frank2021,Mejuto2023a,Lee2023a,Lee2023b}.
However, all these models had exclusively local interactions, simplifying the embedding approximation.
To investigate the applicability of this method to correlated molecular systems, it is imperative to ascertain how to reliably recover non-local interactions, as they are paramount to describing from complex catalytic centres, to simple bond breaking.

In this work, we aim to perform such a study, and propose a gGut based approximation that captures non-local electronic correlation.
After a brief heuristic summary of the main ingredients in the gGut framework, we introduce our approximation and a summary of its algorithmic structure.
We exemplify the capabilities and limitations of this approach on two toy models, the Hubbard dimer at half-filling and the H$_2$ molecule in minimal basis.
We then proceed to studying its performance on two simple yet non-trivial dissociation scenarios: the H$_2$ molecule in the cc-pvDz basis, and the H$_6$ ring in minimal basis.
We conclude the paper with a discussion of how this initial approximation can be enhanced in future steps towards a more rigorous inclusion of non-local correlations within quantum embedding.

\section{Embedding with Auxiliary Quasiparticles - Ghost Gutzwiller}
\begin{figure}
    \centering
     \includegraphics[width=0.5\textwidth]{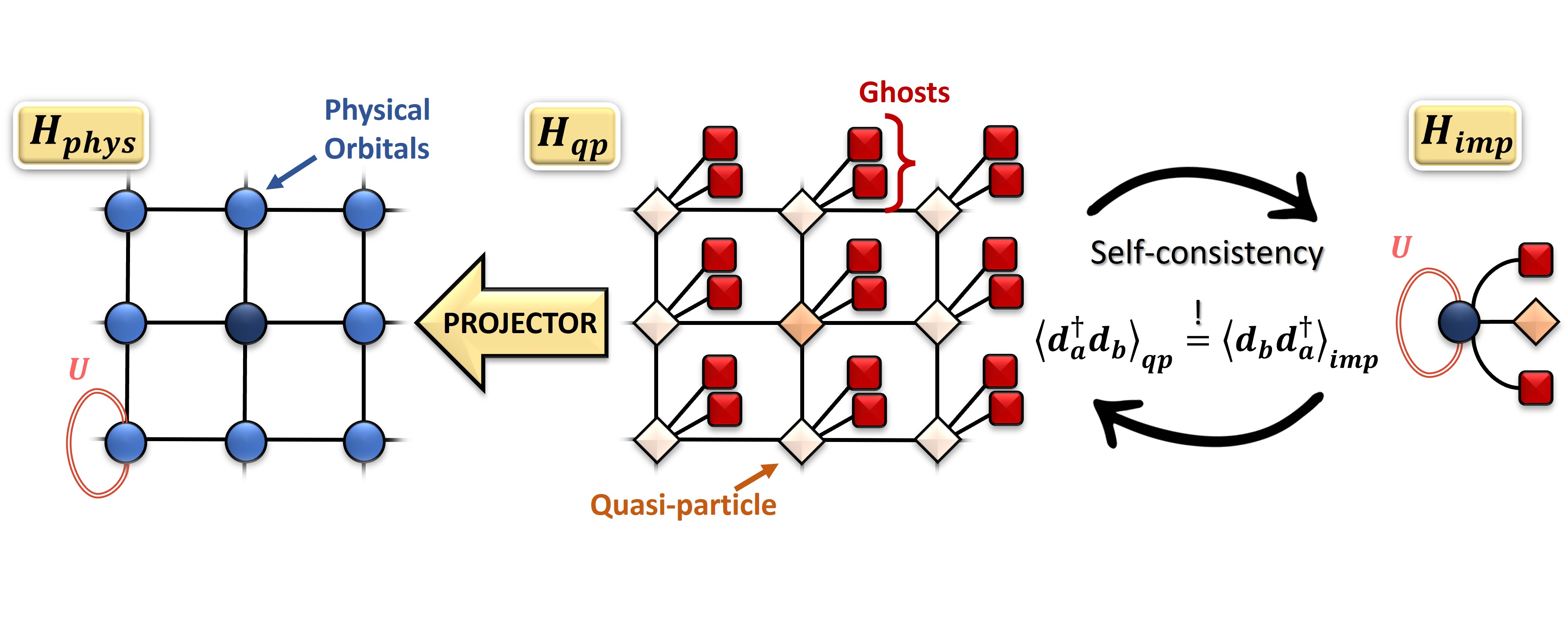}
    \caption{Schematic representation of the Gutzwiller and ghost Gutzwiller Ansatz. Gutzwiller corresponds to the limit with no ghost orbitals. Local interactions are marked with the symbol $U$. See text for details.}
    \label{fig:gGut_schematic}
\end{figure}

This section briefly discusses the Gutzwiller and ghost Gutzwiller approaches from a heuristic point of view.
For detailed derivations of the formalism, as developed for models with exclusively local interactions, we refer to the existing literature~\cite{Bunemann1998,Bunemann2005,Fabrizio2007,lanata2015}. 
Here we provide a narrative description of what the method can provide, as well as the main equations involved.
Finally we introduce an approach to recover the effect of non-local interactions in the gGut framework. 

Briefly, both Gutzwiller and gGut are variational Ans\"atze in which the test wave function is composed of a single Slater determinant and a projection operator. This is a common structure, which can be optimized with Monte Carlo sampling, e.g., in terms of Jastrow factors~\cite{Casula2003, Casula2004,Neuscamman2012,becca2017,Raghav2023}, or other minimization strategies~\cite{Ye2019,Ye2022}.
However, instead of performing the variational optimization exactly, here we apply an approximation based on spatial locality to formulate it instead as an embedding problem~\cite{lanata2015}.
This comes at the price of reducing the quantitative accuracy, but provides with a computationally flexible and comparatively inexpensive method, which has been shown to give direct access to qualitatively faithful spectral information.

\subsection{Heuristic of the Gutzwiller Approximation And the Role of Ghosts}
The Gutzwiller variational wave function $\ket{\Psi_G}$ is composed of two main ingredients: a Slater determinant $\ket{\Psi_{qp}}$, and a projection operator $P$, both of which are optimized to minimize the ground state energy of a physical Hamiltonian of interest $H_{phys}$,

\beal
    \ket{\Psi_G} &= P \ket{\Psi_{qp}},\\ E_0^G &=\min_{P,\Psi_{qp}}\frac{\braket{\Psi_G|H_{phys}|\Psi_G}}{\braket{\Psi_G|\Psi_G}}.
    \label{eq:GutAnsatz}
\eal

The Slater determinant is obtained as the ground state of an effective, one-body model for the system of interest, the quasi-particle Hamiltonian $H_{qp}$, defined below.
Being a one-body model, it can capture correlation effects primarily through a renormalization of the one-body Hamiltonian terms, and possibly by adding an auxiliary single particle potential.
Both physical and quasi-particle Hamiltonians are schematically represented in Fig.~\ref{fig:gGut_schematic}, where physical orbitals are shown as circles and quasi-particle ones as diamonds/squares.

On the other hand, the projector $P$ is a map between the quasi-particle and the physical Hilbert spaces, and thus in general contains a number of variational parameters which is exponentially large in the system size.
Performing an energy optimization over such a number of parameters is a daunting task, computationally expensive and prone to local minima.
We alleviate this complexity by introducing the Gutzwiller approximation, essentially a large coordination number approximation which becomes exact in the limit of infinite dimensions.
Within this local approximation, the projector only acts within a restricted set of physical and quasi-particle orbitals at a time.
In essence, we define a fragmentation in the system, and the projector splits into smaller sub-projectors $P = \prod_I P_I$, which only map onto electronic configurations local to each fragment $I$.
This reduces the number of variational parameters, and further allows us to evaluate the expectation values that enter in Eq.~\eqref{eq:GutAnsatz} analytically~\cite{Fabrizio2007}.
Indeed, we can split the energy expectation value in Eq.~\eqref{eq:GutAnsatz} into terms local to each fragment, and the non-local contributions of the Hamiltonian $H^{phys} = \sum_I H^{loc}_I + H^{latt}$

\beal
    \braket{\Psi_G|H_{phys}|\Psi_G} &= \sum_I\braket{\Psi_{qp}|P^\dagger H^{loc}_{I} P^\dagga|\Psi_{qp}}+\braket{\Psi_{qp}|P^\dagger H^{latt} P^\dagga|\Psi_{qp}},\\
    &= \sum_I \braket{\Psi_{qp}| P^\dagger_I H^{loc}_I P^\dagga_I|\Psi_{qp}} + \braket{\Psi_{qp}|H_{qp}|\Psi_{qp}},
    \label{eq:ProjExpVal}
\eal

where in the last line we have introduced $H_{qp}$, such that it fulfills $\braket{\Psi_{qp}|H_{qp}|\Psi_{qp}} = \braket{\Psi_{qp}|P^\dagger H^{latt} P^\dagga|\Psi_{qp}}$.
Now, within the Gutzwiller approximation and assuming that $H^{latt}$ contains no interactions, $H_{qp}$ can be derived~\cite{Bunemann2005} by performing the following substitution on the physical creation/annihilation operators $c^\dagger_{\alpha}$

\begin{equation}
    P^\dagger_I c^\dagger_{\alpha_I} P^\dagga_I \rightarrow \sum_{a_I} R^{I,\dagger}_{a_I\alpha_I}\ d^\dagger_{a_I},
    \label{eq:Req}
\end{equation}

which essentially defines the quasi-particle orbitals $\left\{d^\dagger_{a_I}\right\}$ in terms of renormalization matrices $\vec{R} = \left\{R^I\right\}$.
We will henceforth use greek indices and operators $c^\dagger_{\alpha_I}$ to refer to physical orbitals, and latin indices $a,b$ and operators $d^\dagger_{a_I}$ for quasi-particle orbitals.
Further, we employ composite indices $x_I$, where capital lettes $I,J$ identify fragments and small case letters identify orbitals.

Even after this simplification, one would need to variationally optimize the parameters defining the local projectors $P_I$, which are still exponential in the local Hilbert space sizes.
Within the already established high coordination number limit, this final optimization can be exactly substituted by finding the ground state of an auxiliary impurity Hamiltonian $H^{imp}_I$ for each sub-projector, shown in the rightmost part of Fig.~\ref{fig:gGut_schematic}.
Like in DMET, in the Gutzwiller method these impurity models have exactly as many bath orbitals as impurity orbitals.
Thus, within the Gutzwiller approximation, it is possible to formulate the variational Ansatz as an embedding-like approach in which local fragments of the quasi-particle Hamitlonian are mapped into impurity models.
The parameters of this mapping are fixed by a self-consistent condition, imposing that the local one-body reduced density matrices (1-RDM) of the quasi-particle Hamiltonian match the bath 1-RDMs of the the impurity models, as shown in Fig.~\ref{fig:gGut_schematic}.
This condition is imposed by adding local one-body potentials $\vec{\lambda}=\left\{\lambda_I\right\}$ to $H_{qp}$, working essentially as Lagrange multipliers.
While the Gutzwiller approximation technically makes the calculation non-variational, except in the infinite-dimensional limit, most if not all previous studies seem to find a nonetheless variational behavior of the converged energy~\cite{lanata2015,Lanata2017,Mejuto2023a,Lee2023a,Lee2023b}.

Once the self-consistency is reached, one obtains a quasi-particle and, potentially several, impurity Hamiltonians which represent the correlated system studied within the Gutzwiller approximation.
Local (fragment) quantities can be evaluated from the impurity models, while non-local information such as momentum dependent energy dispersions are accessed through the quasi-particle Hamiltonian.
Remarkably, this embedding gives qualitatively accurate spectral functions despite spectral information never entering the self-consistency explicitly.
Further, being effectively variational, it can be used to compute forces, a direction which is yet to be fully exploited.

\begin{figure}
    \centering
    \includegraphics[width=0.5\textwidth]{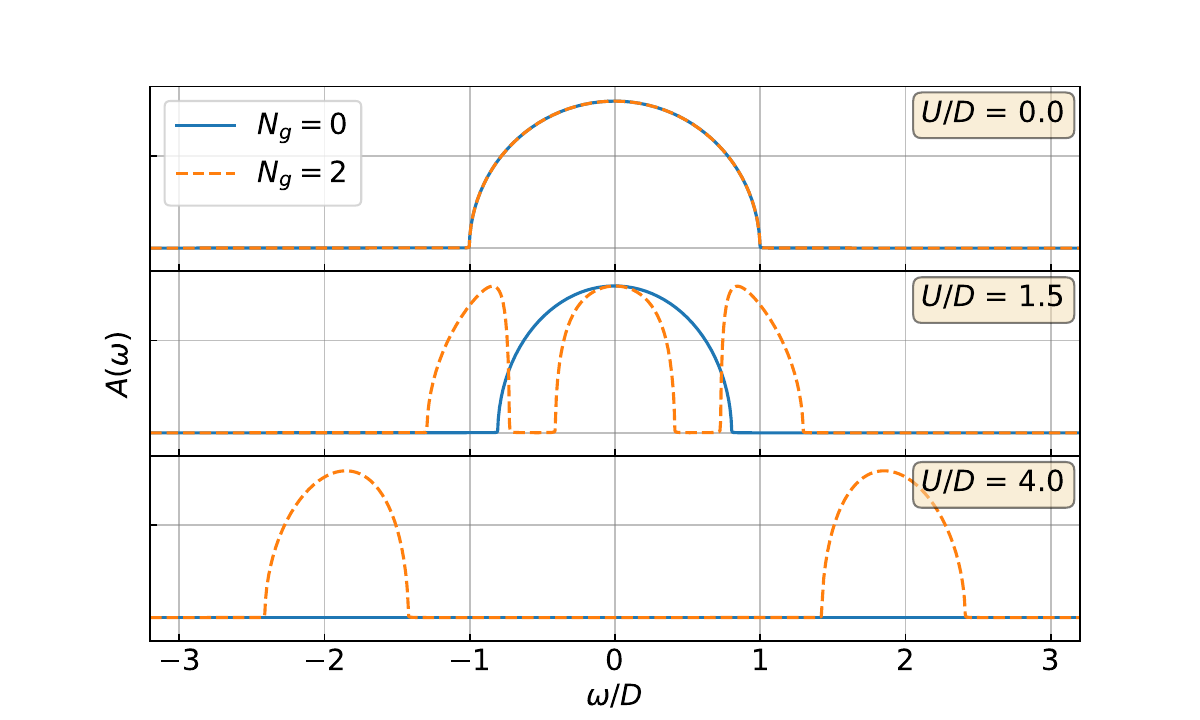}
    \caption{Example of the effect of ghosts when capturing correlated spectra. Spectral function $A(\omega)$ for the one-band Hubbard model at half-filling in the Bethe lattice for 3 different interactions in Gutzwiller ($N_g = 0$) and ghost Gutzwiller ($N_g = 2$) approximations. See text for details.}
    \label{fig:ghosts_example}
\end{figure}

The Gutzwiller approximation has seen great success in the description of strongly correlated phenomena.
Famously, it provided the first picture of the paramagnetic Mott transition within the so-called Brinkman-Rice scenario~\cite{Gutzwiller1963,Gutzwiller1965}.
This can be exemplified by solving the one-band Hubbard model at half-filling in the Bethe lattice, where the Gutzwiller approximation to the expectation values is exact.
Here the fragmentation of the system corresponds to a single site/orbital per fragment, with all sub-projectors and impurity models being equal by space translational invariance.
Fig.~\ref{fig:ghosts_example} shows the resulting spectral function for different interaction strenghts as blue curves.
The onset of Mott localization is shown in a progressive narrowing of the metallic band at zero frequency, and a first order transition to a gapped phase beyond a critical interaction strength.
However, this picture is far from perfect: indeed, the Brinkman-Rice scenario sees a metal to insulator transition with no symmetry breaking, but it does so by losing all spectral weight in the insulating phase (see blue curve in the lower panel of Fig.~\ref{fig:ghosts_example}).
In other words, it misses the high-energy, incoherent Hubbard bands.
From the embedding picture of the Gutzwiller Ansatz, the explanation for this shortcoming is rather simple: since the impurity model has a single impurity orbital, it only has a single bath.
This single bath cannot capture at the same time the metallic band at zero frequency as well as the high-energy Hubbard bands.
Consequently, it is possible to address this limitation of the Gutzwiller Ansatz by adding more bath orbitals to the impurity model: this is the role of the ghosts in gGut.
Ghosts are auxiliary orbitals added to the quasi-particle Hamiltonian (see red squares in Fig.~\ref{fig:gGut_schematic}), which have to be projected out by the projector operator.
They enhance the variational flexibility of the Ansatz, while simultaneously increasing the number of baths into the impurity model.
This simple modification of the method allows capturing the elusive Hubbard bands, see orange curves in Fig.~\ref{fig:ghosts_example}.
Moreover, we have shown how in the multi-orbital settings ghosts allow to correctly capture orbital selective Mott physics, crystal-field mediated insulator-insulator transitions, and Hund's metallicity~\cite{Mejuto2023a}.
It is worth emphasizing that this qualitatively accurate spectral information is obtained from an embedding method which is formulated around the static 1-RDM exclusively.

All previous studies of the gGut Ansatz were performed in systems presenting purely local interactions, but its variational structure does not pose any real limitation \emph{a priori} to apply it on systems with non-local interactions.
Before proposing a way to include these non-local effects, we will briefly review the algorithmic formulation of the approach in lattice models with local interactions, to introduce the fundamental equations.

\subsection{The ghost Gutzwiller Approach for Lattices}
As described above, there are three Hamiltonians entering a gGut calculation: the physical, quasi-particle, and impurity Hamiltonians.
In the lattice setting, the physical Hamiltonian is often assumed to have exclusively local interactions, and thus is written as:

\beal
        H_{phys} &= H^{latt} + \sum_I H^{loc}_{I}, \\
        H^{latt} &= \sum_{I\neq J}\sum_{\alpha_I\beta_J}\, t_{\alpha_I\beta_J}\, c^\dagger_{\alpha_I}\,c^\dagga_{\beta_J} \\
        H^{loc}_{I} &= \sum_{\alpha_I\beta_I}\, t_{\alpha_I\beta_I}\,c^\dagger_{\alpha_I}\,c^\dagga_{\beta_I}\\
        &+\frac{1}{2}\sum_{\alpha_I\beta_I\,\gamma_I\delta_I}\, U_{\alpha_I\beta_I\,\gamma_I\delta_I}\,c^\dagger_{\alpha_I}\,c^\dagger_{\gamma_I}\,
        c^\dagga_{\delta_I}\,c^\dagga_{\beta_I}\,. 
        \label{eq:PhysHamil}
\eal

where we distinguish between local Hamiltonians $H^{loc}_I$ including all terms involving a single site $I$, and a lattice Hamiltonian $H^{latt}$ including the one-body terms between different lattice sites $I,J$.
When performing a gGut calculation on this $H_{phys}$, a natural fragmentation choice corresponds to each site $I$ belonging to its own fragment, and hence the full projector operator is split as a product of local single-site projectors, which are equal in presence of space translation symmetry.
The quasi-particle Hamiltonian will keep the non-local terms, possibly renormalize them with a set of local matrices $\vec{R}$, and add local one-body potentials $\vec{\lambda}$ to model local correlations.

\begin{equation}
    H_{qp} = \sum_{I\neq J}\sum_{a_Ib_J}\left(\sum_{\alpha_I\beta_J}\, R^{I,\dagger}_{a_I\alpha_I}\, t_{\alpha_I\beta_J} \,R^{J\dagga}_{\beta_J b_J}-\delta_{IJ}\lambda^{I}_{a_Ib_I}\right)\ d^\dagger_{a_I}\,d^\dagga_{b_I}.
    \label{eq:qp-Hamil}
\end{equation}

In the normal Gutzwiller approximation, there are as many quasi-particle orbitals as physical ones, whereas in gGut there are more, since one adds also the ghosts.
The matrices $\vec{R}$ and $\vec{\lambda}$ are the self-consistent parameters of the method, since they are the variational parameters defining of the Slater determinant in Eq.~\eqref{eq:GutAnsatz}.
In the presence of space translational invariance, each $R^I$ and $\lambda^I$ are assumed to be equal.
Meanwhile, the variational parameters of the projector are determined by finding the ground state of the impurity models, which carry the local Hamiltonians $H^{loc}_I$.
These impurity models correspond to each isolated fragment of the physical system hybridized with a bath.
This bath derives from the local (fragment) space of the quasi-particle Hamiltonian.

\beal
    H^{imp}_I &= H^{loc}_I + \sum_{\alpha_Ia_I}\,\Big(V^I_{\alpha_Ia_I} d^\dagger_{a_I}\, c^\dagga_{\alpha_I} + \mathrm{h.c.}\Big) -\sum_{a_Ib_I}\, \lambda_{a_Ib_I}^{I,c}\, d^\dagger_{a_I}\,d^\dagga_{b_I}.
    \label{eq:imp-H}
\eal

Note that in these impurity models, there are as many impurity orbitals as physical orbitals in the corresponding fragment in $H_{phys}$, and as many bath orbitals as quasi-particle orbitals in the corresponding fragment in $H_{qp}$. 
Again, in the presence of space translational symmetry, all these impurity models are equal, and hence only one needs to be solved in any given iteration. 
The exact relations between ($\lambda^{I,c}$, $V^I$) and the quasi-particle Hamiltonian $H_{qp}$ are given by

\begin{equation}
    \sqrt{\Delta^{II,qp}\big(\mathbb{I}-\Delta^{II,qp}\big)\;}\cdot V^I = \sum_{J\neq I}\Delta^{IJ,qp}\cdot R^{I,\dagger}\cdot t_{IJ},
    \label{eq:V}
\end{equation}

and

\begin{equation}
    \lambda^{I,c}_{a_Ib_I} = -\lambda^{I}_{a_Ib_I}+\left\{\frac{\partial}{\partial \Delta^{qp}_{a_Ib_I}}
    \left[R^I\cdot \sqrt{\Delta^{II,qp}(\mathbb{I}-\Delta^{II,qp})\;}\cdot V^I \right]+\mathrm{h.c.}\right\}
    \label{eq:lambd_c}
\end{equation}

where we have introduced the quasi-particle 1-RDM $\Delta^{qp}_{a_Ib_J} = \langle d^\dagger_{a_I}d^\dagga_{b_J}\rangle_{qp}$, $\Delta^{IJ,qp}$ refers to the block of the quasi-particle 1-RDM corresponding to fragments $I,J$, and similarly for $t_{IJ}$ for the one-body Hamiltonian components. 
The self-consistency condition of the gGut embedding is formulated in terms of the quasiparticle and impurity model 1-RDMs as:

\begin{equation}
    \langle\, d^\dagga_{b_I}\,d^\dagger_{a_I}\,\rangle_{imp} = \delta_{a_Ib_I}-\Delta^{I,imp}_{a_Ib_I} \overset{!}{=} \Delta^{qp}_{a_Ib_I} = \langle\, d^\dagger_{a_I}\, d^\dagga_{b_I} \,\rangle_{qp}\,.
    \label{eq:scf}
\end{equation}

Since the self-consistency only involves the static 1-RDM, the full simulation is computationally less expensive than embedding approaches requiring the impurity GF, like DMFT.
The self-consistency condition in Eq.~\eqref{eq:scf} is typically enforced iteratively.
At the $\ell$-th iteration we run through the above equations, and finally propose a new $\vec{R}$ and $\Delta^{qp}$ as 

\begin{equation}
    \begin{split}
        \Delta^{II,qp,\ell+1} &= \mathbb{I} - \Delta^{I,imp,\ell}_{bath-bath}, \\
        R^{I,\ell+1}\cdot\sqrt{\Delta^{II,qp,\ell+1}(\mathbb{I}-\Delta^{II,qp,\ell+1})} &= \Delta^{I,imp,\ell,t}_{bath-imp}.
    \end{split}
    \label{eq:scf_cond}
\end{equation}

Where $\Delta^{I,imp}_{bath-imp}$ denotes the off-diagonal block of the $I$-th impurity model RDM between the impurity and bath orbitals, and $\Delta^{I,imp}_{bath-bath}$ the corresponding bath-bath block.
Obtaining a new $\vec{\lambda}$ in the following iteration implies thus formally a fitting problem, to enforce that the quasi-particle Hamiltonian in Eq.~\eqref{eq:qp-Hamil} has the 1-RDM obtained from Eq.~\eqref{eq:scf_cond}.
This fitting step can be substituted by using the equation for $\lambda^c$ in Eq.~\eqref{eq:lambd_c} to propose a new $\lambda$, which at self-consistency typically leads to a $H_{qp}$ with the right 1-RDM.

Once the self-consistency is reached, we obtain a set of $\vec{R}$ and $\vec{\lambda}$ defining the variationally optimal gGut Ansatz.
From this, observables of interest can be extracted, such as the ground state energy $E_0^G$ or the spectral function $A(\omega)$.
The ground state energy follows

\begin{equation}
    E_0^G = \sum_I E_{imp}^I + \sum_{I\neq J}\sum_{a_Ib_J} \left(\sum_{\alpha_I\beta_J}\, R^{I,\dagger}_{a_I\alpha_I}\, t_{\alpha_I\beta_J} \,R^{J\dagga}_{\beta_J b_J}\right)\ \Delta^{qp}_{a_Ib_J},
    \label{eq:GutzEnergy}
\end{equation}

where the first term is the sum of the ground state energies $E^I_{imp}$ of the impurity models, while the second term is the ground state energy of $H_{qp}$ minus the contribution from the on-site potentials $\vec{\lambda}$. 
The spectral function is related to the imaginary part of the Green's function as $A(\omega) = -\frac{1}{\pi}\Im\left[\mathrm{Tr}G(\omega)\right]$, and the Green's function can be obtained as

\begin{equation}
    G_{\alpha_I\beta_J}(\omega) = \sum_{ab}R^{I\dagga}_{\alpha_Ia_J}\,\left[\frac{1}{(\omega+i0^+)\mathbb{I} - H_{qp}}\right]_{a_Ib_J}\,R^{J,\dagger}_{b_J\beta_J}.
    \label{eq:gGut_GF}
\end{equation}

Essentially, one has to compute the Green's function of the quasi-particle Hamiltonian (the resolvent on the right hand side of Eq.~\eqref{eq:gGut_GF}), and project it back to the physical space using the $\vec{R}$ matrices.

\subsection{The ghost Gutzwiller Approach for Molecules}
In order to apply the gGut formalism to molecular systems, we have to drop the assumption that the physical Hamiltonian includes only local interactions. 
Regardless of what fragmentation we propose for the Gutzwiller embedding, there will be in general interactions beyond any fragment Hilbert space.
Given the comparative lack of symmetries between most molecular systems and crystalline solids, we will also generalize the structure of the projection operator, allowing for different local projectors for each fragment.
Finally, unlike in the lattice model case above, real systems typically have uncorrelated electrons, which are in orbitals or bands well above or below the Fermi level.
For this reason, we want to allow for a set of orbitals, henceforth the ``uncorrelated space'', which will not be acted upon by any projector.
For this orbital subset, we will reserve the index $Y$.
The molecular Hamiltonian can hence be divided into three components

\beal
    H_{phys} &= \sum_I H^{loc}_{I} + H^{loc}_Y + H^{hyb},\\
    H^{loc}_{I} &= \sum_{\alpha_I\beta_I,\sigma}t_{\alpha_I\beta_I}\ c^\dagger_{\alpha_I\sigma}c^\dagga_{\beta_I\sigma}\\
    &+\frac{1}{2}\sum_{\substack{\alpha_I\beta_I\gamma_I\delta_I\\ \sigma\sigma'}}U_{\alpha_I\beta_I\,\gamma_I\delta_I}\ c^\dagger_{\alpha_I\sigma}c^\dagger_{\gamma_I\sigma'}c^\dagga_{\delta_I\sigma'}c^\dagga_{\beta_I\sigma},\\
    H^{loc}_Y &= \sum_{\alpha_Y\beta_Y,\sigma}t_{\alpha_Y\beta_Y}\ c^\dagger_{\alpha_Y\sigma}c^\dagga_{\beta_Y\sigma}\\
    &+\frac{1}{2}\sum_{\substack{\alpha_Y\beta_Y\gamma_Y\delta_Y\\ \sigma\sigma'}}U_{\alpha_Y\beta_Y\,\gamma_Y\delta_Y}\ c^\dagger_{\alpha_Y\sigma}c^\dagger_{\gamma_Y\sigma'}c^\dagga_{\delta_Y\sigma'}c^\dagga_{\beta_Y\sigma},\\
    H^{hyb} &= \sum_{I\neq J}H^{hyb}_{IJ} + \sum_I H^{hyb}_{IY},
    \label{eq:Mol_frag}
\eal

where $H^{hyb}_{IJ}$ collects the one- and two-body terms which include orbitals from at least two distinct fragments $I,J$, and $H^{hyb}_{IY}$ collects the terms including orbitals only from fragment $I$ and at least one uncorrelated orbital.
Note that in our lattice example above, $H^{loc}_Y$ would be exactly zero, $H^{Hyb}$ would be exclusively composed of hoppings between different lattice sites, and all correlated Hamiltonians $H^{loc}_I$ would be equal in the presence of space translational invariance.

Now, within this fragmentation, we will propose a variational Ansatz based on the following projector

\begin{equation}
    \ket{\Psi_G} = P\ket{\Psi_{qp}} = \prod_{I} P_I \ket{\Psi_{qp}},
    \label{eq:Mol_Proj}
\end{equation}

where each local projector $P_I$ may be different from each other, and there is no projection operator acting on the uncorrelated orbitals (which can be seen as $P_Y = \mathbb{I}$).
Because of this, we cannot add ghosts to these orbitals.
This is not a severe approximation, since the underlying Slater determinant can be exact for uncorrelated electrons.

In principle, it is possible to apply the standard derivation of the Gutzwiller formalism to the Hamiltonian in Eq.~\eqref{eq:Mol_frag}, the same way as for lattice models, to arrive at an embedding description.
This can be done by writing out the expectation values of the different Hamiltonian terms using the Ansatz in Eq.~\eqref{eq:Mol_Proj}, and invoking the Gutzwiller, i.e., large coordination number, approximation.
Since each correlated fragment has a different projector, each will generate a different impurity model, which will have to be solved explicitly in every iteration.
Besides this, the main differences with the case of local interactions will be that (i) the impurity model will include interactions between the impurity and bath orbitals and (ii) the quasi-particle Hamiltonian will feature all the interaction terms in $H^{loc}_Y$ and $H^{hyb}$.
The interactions in $H^{loc}_Y$ will appear as-is, while those in $H^{hyb}$ will be renormalized by tensors $\chi$  arising from the Gutzwiller approximation analogously to the $R$ tensors in Eq.~\eqref{eq:Req}, e.g., as

\begin{equation}
    P^\dagger_I c^\dagger_{\alpha_I} c^\dagga_{\beta_I}P^\dagga_I \rightarrow \sum_{a_Ib_I} \chi^{I}_{a_Ib_I\, \alpha_I\beta_I}\ d^\dagger_{a_I} d^\dagga_{b_I}.
    \label{eq:Xeq}
\end{equation}

While this would correspond to the most rigorous application of the gGut Ansatz to the molecular Hamiltonian, and hence would account for non-local interactions in the most accurate way, the proliferation of additional self-consistent parameters $\chi$ will also greatly increase the complexity of the self-consistency convergence.
To gauge how well this type of variational embedding may capture non-local correlations in molecular systems, and whether this type of strategy offers any potential advantage, it is advisable to instead start from a simplified approximation to the Ansatz.
This is the goal of the present work, and the simplification will be performed by decoupling the interactions in $H^{loc}_Y$ and $H^{hyb}$ in mean-field.

\subsection{Mean-Field decoupling of Non-Local Interactions}
The simplest approximation to be performed on the non-local interactions is a mean-field decoupling following

\beal
    c^{\dagger}_{\alpha\sigma}c^\dagger_{\gamma\sigma'}c^\dagga_{\delta\sigma'}&c^\dagga_{\beta\sigma} \rightarrow\  c^\dagger_{\alpha\sigma}c^\dagga_{\beta\sigma}\langle c^\dagger_{\gamma\sigma'}c_{\delta\sigma'}\rangle+c^\dagger_{\gamma\sigma'}c^\dagga_{\delta\sigma'}\langle c^\dagger_{\alpha\sigma}c_{\beta\sigma}\rangle\\
    &-\delta_{\sigma\sigma'}\left[c^\dagger_{\alpha\sigma}c^\dagga_{\delta\sigma}\langle c^\dagger_{\gamma\sigma}c_{\beta\sigma}\rangle+c^\dagger_{\gamma\sigma}c^\dagga_{\beta\sigma}\langle c^\dagger_{\alpha\sigma}c_{\delta\sigma}\rangle\right]\\
    &-\langle c^\dagger_{\alpha\sigma}c^\dagga_{\beta\sigma}\rangle\langle c^\dagger_{\gamma\sigma'}c_{\delta\sigma'}\rangle+\delta_{\sigma\sigma'}\langle c^\dagger_{\alpha\sigma}c^\dagga_{\delta\sigma}\rangle\langle c^\dagger_{\gamma\sigma}c_{\beta\sigma}\rangle.
    \label{eq:mf_decoupling}
\eal

We will assume for simplicity a restricted (i.e. spin independent) mean field, such that $\langle c^\dagger_{\alpha\sigma}c^\dagga_{\beta\sigma}\rangle = \langle c^\dagger_{\alpha}c^\dagga_{\beta}\rangle\equiv \Delta^{mf}_{\alpha\beta}$.
The idea is to perform the mean-field decoupling directly on the molecular Hamiltonian, before applying the gGut Ansatz.
This decoupling is performed on all non-local interaction terms, such that the one-body terms in Eq.~\eqref{eq:Mol_frag} become

\beal
    \tilde{t}_{\alpha\beta} &= t_{\alpha\beta}+\sum_{I,J} u^{IJ}_{\alpha\beta} + \sum_{I}\left[u^{IY}_{\alpha\beta}+u^{YI}_{\alpha\beta}\right]+u^{YY}_{\alpha\beta}-u^{DC}_{\alpha\beta},\\
    u^{AB}_{\alpha\beta} &= \sum_{\substack{\gamma_A\in A\\ \delta_B\in B}}\left[2U_{\alpha\beta\,\gamma_A\delta_B}-U_{\alpha\delta_B\,\gamma_A\beta}\right]\Delta^{mf}_{\gamma_A\delta_B},\\
    u^{DC}_{\alpha\beta} &= \begin{cases} 
      u^{II}_{\alpha\beta} & \mathrm{if}\ \alpha\ \mathrm{and}\ \beta\ \mathrm{both\ in\ the\ same\ fragment}\ I \\
      0 & \mathrm{otherwise} 
   \end{cases},
    \label{eq:MF-decoup-Hamil}
\eal

where the final term $u^{DC}$ is a double counting correction, since local fragment interactions are not mean-field decoupled.
Note that, in a fragmentation involving multiple fragments, this type of mean-field decoupling does not correspond to an expectation value on the original Hamiltonian in Eq.~\eqref{eq:Mol_frag}.
Therefore, despite the gGut Ansatz being still variational up to the local approximation, this complete scheme of mean-field + gGut (mf+gGut) loses the variational property.

After applying the mean-field decoupling in Eq.~\eqref{eq:MF-decoup-Hamil}, the molecular Hamiltonian becomes an effective impurity-model-like Hamiltonian, composed exclusively of locally interacting fragments, and possibly a non-interacting, uncorrelated subspace, following:

\beal
    \tilde{H}_{phys} &= \sum_I \tilde{H}^{loc}_{I} + \tilde{H}^{loc}_Y + \tilde{H}^{hyb},\\
    \tilde{H}^{loc}_{I} &= \sum_{\alpha_I\beta_I,\sigma}\tilde{t}_{\alpha_I\beta_I}\ c^\dagger_{\alpha_I\sigma}c^\dagga_{\beta_I\sigma}\\
    &+\frac{1}{2}\sum_{\substack{\alpha_I\beta_I\gamma_I\delta_I\\ \sigma\sigma'}}U_{\alpha_I\beta_I\,\gamma_I\delta_I}\ c^\dagger_{\alpha_I\sigma}c^\dagger_{\gamma_I\sigma'}c^\dagga_{\delta_I\sigma'}c^\dagga_{\beta_I\sigma},\\
    \tilde{H}^{loc}_Y &= \sum_{\alpha_Y\beta_Y,\sigma}\tilde{t}_{\alpha_Y\beta_Y}\ c^\dagger_{\alpha_Y\sigma}c^\dagga_{\beta_Y\sigma},\\
    \tilde{H}^{hyb} &= \sum_{I\neq J}\sum_{\alpha_I\beta_J,\sigma}\tilde{t}_{\alpha_I\beta_J}\ c^\dagger_{\alpha_I\sigma}c^\dagga_{\beta_J\sigma}\\
    &+\sum_{I}\sum_{\alpha_I\beta_X,\sigma}\left(\tilde{t}_{\alpha_I\beta_X}\ c^\dagger_{\alpha_I\sigma}c^\dagga_{\beta_X\sigma}+\mathrm{h.c.}\right),
    \label{eq:Mol_frag_mf}
\eal

This Hamiltonian can then be treated with the gGut Ansatz in exactly the same way as the lattice Hamiltonian in the previous subsection, since the only explicit interactions it presents are fully localized to the fragments.
The only difference from the calculation described in the previous subsection is that the mean-fields for the non-local interaction decoupling, which enter the effective one-body terms $\tilde{t}$, have to be determined self-consistently with the Gutzwiller solution.
In essence, the expectation values $\Delta^{mf}_{\alpha\beta} = \langle c^\dagger_{\alpha}c^\dagga_{\beta}\rangle$ have to be evaluated within the gGut approximation, namely

\beal
    \Delta^{mf}_{\alpha\beta} = \begin{cases}
        \langle c^\dagger_{\alpha}c^\dagga_{\beta} \rangle_{imp} & \mathrm{if}\ \alpha, \beta\in I \\
        \sum_{ab}\,R^{I,\dagger}_{a\alpha}\,\langle d^\dagger_{a}d^\dagga_{b} \rangle_{qp}\, R^{J\dagga}_{\beta b} & \mathrm{if}\ \alpha\in I, \beta\in J, I\neq J\\
        \sum_{a}\,R^{I,\dagger}_{a\alpha}\,\langle d^\dagger_{a}d^\dagga_{\beta} \rangle_{qp}  & \mathrm{if}\ \alpha \in I, \beta\in X\\
        \langle d^\dagger_{\alpha}d^\dagga_{\beta} \rangle_{qp}  & \mathrm{if}\ \alpha, \beta\in Y\\
        \end{cases}
    \label{eq:mf_in_gGut}
\eal

With this prescription, the mean-field and gGut self-consistencies can be performed alternately: start with some guess mean-field, e.g., from a Hartree-Fock calculation, then run a gGut calculation at fixed mean-field, update the mean-fields with the converged gGut wave function, and repeat until the mean-fields are converged.
The steps in the algorithm are presented in Alg.~\ref{alg:HFgGut}.
One can also consider running this approximation in a ``one-shot'' fashion, i.e., without the outer self-consistency for the mean-field.

\begin{algorithm}
\caption{Mean-Field + ghost Gutzwiller algorithm.}\label{alg:HFgGut}
    \SetKwInOut{Input}{Input}
    \SetKwInOut{Output}{Output}
    \Input{$t,\ U$}
    \Output{$\Delta^{mf},\ \vec{R},\ \vec{\lambda}$}
    $\Delta^{mf} \gets \mathrm{HF}$\ ;\\
    \While{$\Delta^{mf}$ not converged}{
        $\vec{R},\vec{\lambda}\gets \vec{R}_0,\vec{\lambda}_0$\ ;\\
        \While{$\vec{R},\vec{\lambda}$ not converged}{
            $\Delta^{qp}\ \ \ \ \gets H_{qp}\gets\vec{R},\vec{\lambda}$\ \ \ \ Eq.~\eqref{eq:qp-Hamil}\ ;\\
            $\vec{V}\ \ \ \ \ \ \ \gets \Delta^{qp}, \vec{R}$\ \ \ \ \ \ \ \ \ \ \ \ Eq.~\eqref{eq:V}\ ;\\
            $\vec{\lambda^c}\ \ \ \ \ \ \gets \Delta^{qp},\vec{\lambda},\vec{R},\vec{V}$\ \ \ \ Eq.~\eqref{eq:lambd_c}\ ;\\
            $\vec{\Delta}^{imp}\ \ \ \gets \vec{H}_{imp}\gets \vec{V},\vec{\lambda^c}$ Eq.~\eqref{eq:imp-H}\ ;\\
            $\vec{R},\ \Delta^{qp}\gets \vec{\Delta}^{imp}$\ \ \ \ \ \ \ \ \ \ \ \ \ \ Eq.~\eqref{eq:scf_cond}\ ; \\
            $\vec{\lambda}\ \ \ \ \ \ \ \ \gets \Delta^{qp},\vec{\lambda^c},\vec{R},\vec{V}$\ \  Eq.~\eqref{eq:lambd_c}\ ;\\
        }
        $\Delta^{mf}\gets \vec{R},\Delta^{qp}, \vec{\Delta}^{imp}$ Eq.~\eqref{eq:mf_in_gGut}\ ;\\
    }
\end{algorithm}

We have a hierarchy of Gutzwiller-like approximations to treat the electronic correlation in molecular systems: (i) one-shot mf+gGut approximation, (ii) self-consistent mf+gGut and (iii) the rigorous gGut Ansatz without the previous mean-field decoupling.
In the following sections, we will examine the performance of the first two approaches for a series of bond breaking scenarios in small molecular systems, since the description of any bond breaking phenomenon requires capturing non-local correlations correctly.
From the following results, it will become apparent that both one-shot and self-consistent formulations provide with qualitatively equivalent, when not quantitatively identical, descriptions.
This will be useful in calculations for larger systems, in which each gGut self-consistency may require a significant amount of computational time.

\section{Embedding of Toy Models}
To exemplify the capabilities of the mf+gGut Ansatz, and illustrate the features discussed in the previous section, we employ this approximation on two toy models: the Hubbard dimer at half-filling, and the H$_2$ molecule in minimal basis (sto-3g).
We will concentrate on the quality of the ground state energy and spectral function $A(\omega)$ for different fragmentations and numbers of ghosts $N_g$.
Throughout this section, we will use the following short-hand notation to label fragmentations: $(N_{frag}\times N_{orb})$.
This denotes a fragmentation in which there are $N_{frag}$ fragments, each containing $N_{orb}$ orbitals.
When appropriate, the nature of these orbitals will be specified in the text.
For all the calculations in this section, both the one-shot and self-consistent formulation of the mf+gGut approximation give essentially identical results, so only self-consistent data will be reported.

\subsection{The Hubbard Dimer - Role of Multiple Fragments and Ghosts}
\begin{figure}
    \centering
    \includegraphics[width=0.4\textwidth, height=10cm]{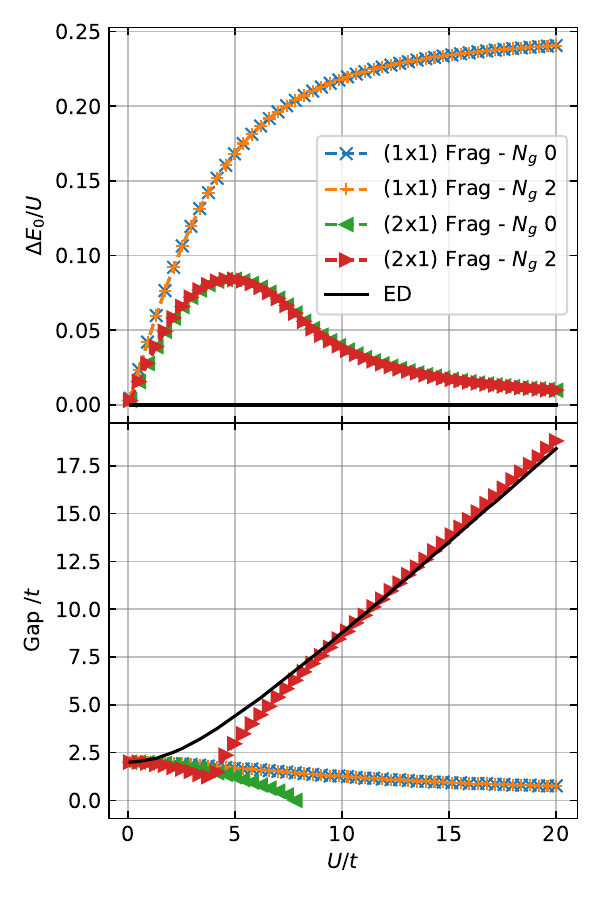}
    \caption{Ghost Gutzwiller calculations on the Hubbard dimer at half-filling, for different interaction strengths $U$ given in units of the hopping $t$. Reported are the error of the ground state energy, divided by $U$ (upper panel) and the gap of the spectral function (lower panel). Results are shown for two different fragmentations, with and without ghosts. See text for details.}
    \label{fig:HubDim}
\end{figure}

Simple though it is, the Hubbard dimer at half-filling already presents some of the main ingredients characterizing strong correlation, which has made it a common benchmark case to study new approximations~\cite{Romaniello2012,Carrascal2015,Mejuto2022b,Riva2022,Giarrusso2023,Riva2023,orlando2023}.
The Hamiltonian reads

\beal
    H_{HD} = &-t\sum_{\sigma}\left(c^\dagger_{1\sigma}c^\dagga_{2\sigma}+\mathrm{h.c.}\right) + U\sum_{I=1,2}n_{I\uparrow}n_{I\downarrow}\\
    &-\frac{U}{2}\sum_{I=1,2,\sigma}n_{I\sigma},
    \label{eq:HD_hamil}
\eal

where $t$ is the hopping between both sites, $U$ the local Hubbard repulsion, and $n_{I\sigma} = c^\dagger_{I\sigma}c^\dagga_{I\sigma}$.
This system is gaped for all interaction strengths $U$, with the gap increasing linearly with $U$ in the large $U$ limit.
A mean-field solution of the problem changes character at $U/t = 2$, the point at which a symmetry broken (unrestricted) Hartree-Fock solution is variationally better than a symmetric (restricted) solution.
In this sense, its behavior as a function of $U/t$ shares many similarities with a bond breaking scenario, transitioning from a delocalized to a localized nature. 
Thus, despite not having explicitly non-local interaction terms in the Hamiltonian, the Hubbard dimer example will allow us to illustrate the role of the embedding fragmentation and number of ghosts.

In Fig.~\ref{fig:HubDim} we present the error in the ground state energy $\Delta E_0$ from different ghost Gutziller Ans\"atze (upper panel), and the gap of the spectral function (lower panel), both as a function of the interaction strength $U/t$.
Since there are only two-orbitals, we can only choose two different, non-trivial, fragmentations: taking one of the two orbitals as a correlated fragment, while leaving the other orbital uncorrelated $(1\times1)$, and correlating both orbitals in different fragments ($2\times1$).
For both fragmentations, we consider calculations with no ghosts, hence pure Gutzwiller Ans\"atze, and simulations with two ghosts for each correlated orbital.
Note that in the $(2\times1)$ simulations, no explicit mean-field decoupling of the interactions is undertaken, although each impurity is only allowed to couple through the one-body $H_{qp}$.

Both the ground state energy error $\Delta E_0$ and the spectral gap in Fig.~\ref{fig:HubDim} show that correlating only one of the two site-orbitals leads to essentially mean-field results: $\Delta E_0$ has an approximate asymptote linear with $U/t$, and the spectral gap has a weak dependence from $U/t$, closing slightly instead of linearly opening up.
This is not surprising, since the uncorrelated orbital remains completely in mean-field in $H_{qp}$.
In restricted mean-field, the Hubbard repulsion becomes a chemical potential shift, and hence electrons in the second orbital cannot localize at large repulsion strengths unless they belong to a correlated fragment, with its own impurity model.
Adding ghosts to the other orbital cannot change this behavior, and indeed the results for the $(1\times1)$ fragmentation with and without ghosts are identical.

Correlating both orbitals in different fragments immediately recovers the right asymptotic behaviour of the ground state energy in the large interaction limit, with a significant error present only around the intermediate interaction regime $U/t\sim 5$.
However, before the introduction of ghosts, the spectral behavior is still qualitatively wrong.
Indeed, the $(2\times1)$ simulations with $N_g = 0$ show a spectral gap that very closely resembles the Brinkman-Rice scenario of the Mott transition discussed above: the gap initially is reduced, and at a critical interaction strength completely closes.
Examining the full spectral function $A(\omega)$ reveals that after this critical interaction ($U/t\sim7.8$), the spectrum is exactly zero, owing to $R$ tending to zero in this limit, exactly as in the treatment of a Hubbard lattice with the Gutzwiller approximation.
Just like in that case, this can be resolved by adding ghosts.
Introducing two ghosts per correlated orbital is enough to recover the right asymptotic behaviour of the gap at large interactions.
There is still a first order transition between this behaviour at large $U/t$, and the low $U/t$ regime, in which the gap still decreases with increasing $U/t$.
In essence, the gGut description of the Hubbard dimer inherits the properties of the gGut model for the Mott transition in the Hubbard lattice, since the underlying impurity models in both cases are essentially equivalent.
This motif will be repeated through the different systems studied in this work.

\subsection{The H$_2$ Molecule in Minimal Basis - Non-Local Approximations}
\begin{figure}
    \centering
    \includegraphics[width=0.5\textwidth]{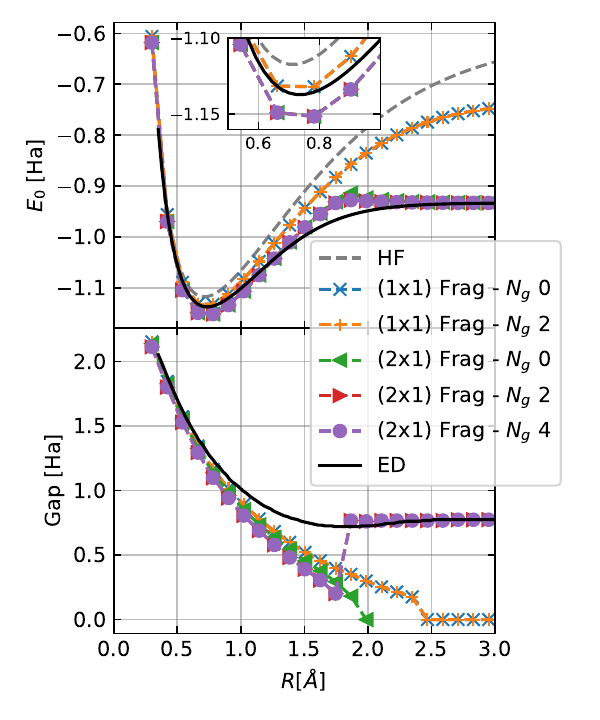}
    \caption{Ghost Gutzwiller calculations on the Hydrogen molecule in sto-3g basis, as a function of intermolecular distance $R$. Reported are the ground state potential energy surface (upper panel) and the gap of the spectral function (lower panel). Results are shown for two different fragmentations, with and without ghosts. See text for details.}
    \label{fig:H2min}
\end{figure}

The H$_2$ molecule in minimal basis (sto-3g) is essentially a Hubbard dimer with dense interaction tensor $U_{\alpha\beta\,\gamma\delta}$.
Hence, we can use it as middle step between toy-models and more reasonable \emph{ab initio} molecular models.
We consider the ground state energy $E_0$ and the spectral gap of the H$_2$ molecule as a function of the inter-atomic distance $R$ for the same fragmentations as in the Hubbard dimer case, now in terms of the two atomic 1s orbitals, and for different numbers of ghosts $N_g$, shown in Fig.~\ref{fig:H2min}.
Note that in this case, all fragmentations involve explicit mean-field decouplings of some interaction terms.
Here, the orbital fragmentations were defined using orthonormal, atomic-like orbitals obtained by rotating the canonical orbitals from restricted HF calculations.

The behaviour of the gGut approximation in the H$_2$ molecular case is completely analogous to that of the Hubbard dimer.
Examining the ground state energy $E_0$, we observe again that correlating just one of the 1s atomic orbitals is only marginally better than Hartree-Fock (HF) in the dissociation regime, while there is a reasonable improvement in the bound region.
Correlating both 1s orbitals separately, the energy recovers the right behaviour at the dissociation limit, and we observe a region of major disagreement somewhere around the Coulson-Fisher point.
Notably, around the bound region the $(2\times1)$ fragmentations report non-variational energies, a consequence of the multi-fragment version of the mean-field decoupling introduced in Eq.~\eqref{eq:Mol_frag_mf}, rather than of the Gutzwiller approximation itself.
In this light, the better agreement of the energy in the $(1\times1)$ fragmentation may be due to error cancellation between the mean-field decoupling in Eq.~\eqref{eq:Mol_frag_mf} and the quasi-particle approximation for one of the 1s orbitals in the Gutzwiller Ansatz.
For either fragmentation, adding ghosts does not seem to significantly change the energy.

Examining the spectral gap in the lower panel of Fig.~\ref{fig:H2min} presents a similar picture as with the Hubbard dimer.
Here, since the exact gap actually closes with increasing inter-atomic distance, the Brinkman-Rice behavior of the gap is at first fortuitously correct for all gGut Ans\"atze.
Nevertheless, the dissociation limit again differenciates between the various fragmentations, and highlights the role of the ghosts.
All $(1\times1)$ simulations, and the $(2\times1)$ one without ghosts, fundamentally fail to capture the spectral gap at large inter-atomic distance, and in fact completely close the gap in the dissociation limit.
The $(1\times1)$ simulations do not lose all spectral weight, while the $(2\times1)$ does, following completely the Brinkman-Rice scenario. 
Once ghosts are included in the $(2\times1)$ fragmentation, upon a first-order phase transition the right gap asymptote is recovered in the dissociation limit.

The analysis of these two toy models shows that the main features of how the Gutzwiller framework models correlation, particularly regarding the role of ghosts in capturing spectral properties, survive in the mf+gGut approximation to non-local interactions.
We can now turn our attention to more complex test cases.

\section{Embedding of Non-Trivial Molecules}
In this section, we employ the mf+gGut Ansatz on two distinct limits of non-trivial bond dissociation: (i) the case of the H$_2$ molecule in cc-pvDz basis, and (ii) the H$_6$ molecular ring in minimal basis (sto-3g).
These correspond to two distinct limits for an embedding treatment.
In the first one, only a small fraction of the orbitals are actually correlated (arguably 2 out of all 10), such that a modest fragmentation leaving most orbitals in the uncorrelated space should reproduce the main low-energy features of the system reliably.
In the second case, all orbitals are actually correlated, so full fragmentations of the system are necessary.
In both cases dissociations, the system undergoes a process of electron localization in which nevertheless the non-local Coulomb interactions are crucial for an accurate description of the energetics, and hence represent stringent tests of the actual reliability of our proposed scheme in the \emph{ab initio} setting.
This type of system has been considered previously for understanding how other embedding approaches fare in systems with non-local interactions~\cite{Lin2011,Knizia2013,Lan2015,Wouters2016,Lee2017}.

We use the same atomic-like, orthonormalized, localized orbitals to define the embedding fragmentations in each case.

\subsection{Diatomic Dissociation - H$_2$ in cc-pvDz basis}

\begin{figure*}
    \centering
    \includegraphics[width=\textwidth]{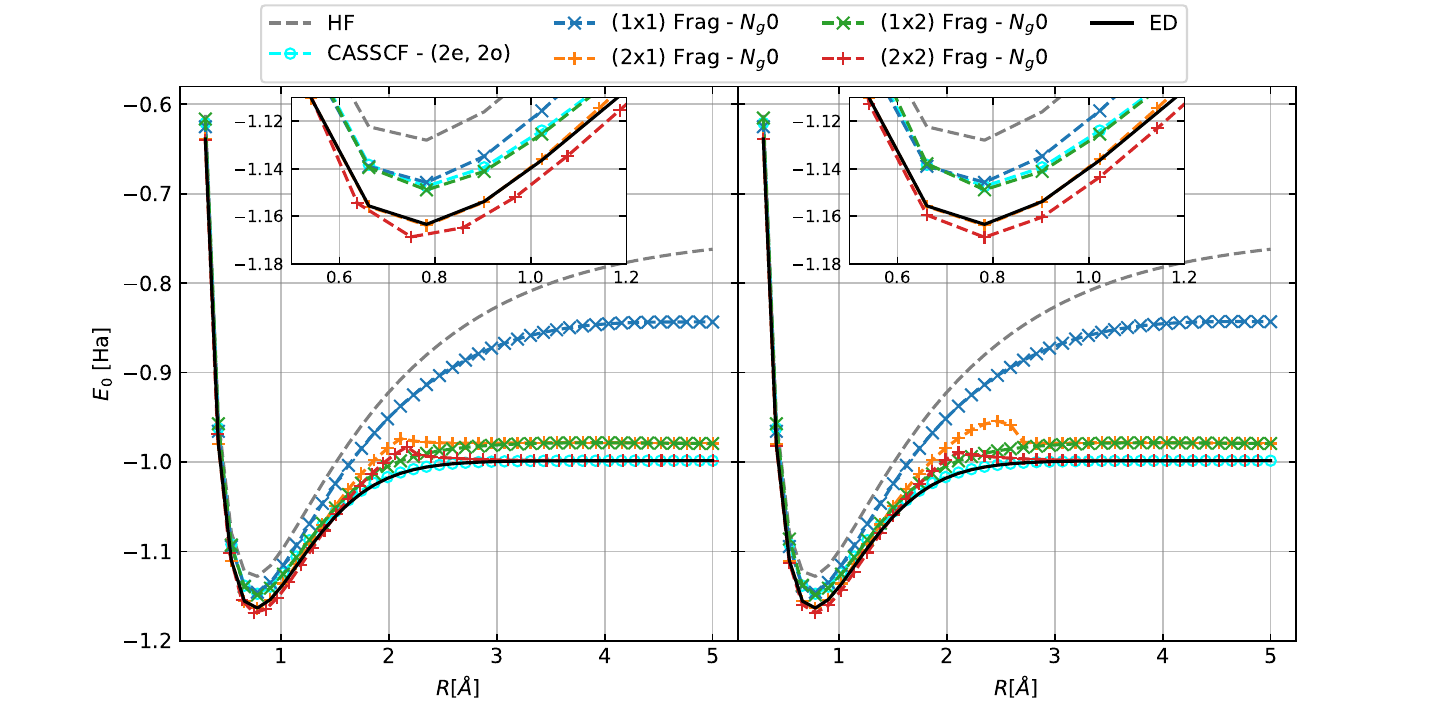}
    \caption{Potential energy surface for the dissociation of the Hydrogen molecule in cc-pvDz basis, within Gutzwiller embedding with different fragmentations. Non-local interactions are decoupled in mean-field, which couples with the Gutzwiller embedding in a one-shot (right panel) and self-consistent (left panel) manner. Different fragmentations are tested, see text for details.}
    \label{fig:H2ccpvdz}
\end{figure*}

The H$_2$ molecule model in cc-pvDz basis consists of 5 orbitals per atom, which can be localized into a 1s, a 2s, and three 2p orbitals.
For the purposes of describing the $\sigma$ bond breaking when separating both H atoms, arguably only 2 orbitals out of the 10 should be necessary: in essence, a bonding/anti-bonding pair.
Consequently, we can compare the effect of different fragmentations on the description of this system with mf+gGut, and use as reference both exact results and a complete active space self-consistent field (CASSCF) calculation with 2 electrons and 2 orbitals (2e, 2o), arguably a static form of embedding.
Further, we will show the differences between performing mf+gGut calculations in the single shot and self-consistent variants.

For the case of H$_2$ in cc-pvDz basis, we will concentrate on the ground state across the dissociation, shown in Fig.~\ref{fig:H2ccpvdz}, and defer an analysis of spectral features for H$_6$ in the following subsection.
Hence, and since the previous results show that the ghosts have little effect on the energy, we will show results for $N_g = 0$ simulations.
We consider the following fragmentations:

\begin{enumerate}
    \item $(1\times1)$: Single fragment composed of the 1s orbital of one of the H atoms.
    \item $(2\times1)$: Two fragments, each with the 1s orbital of each atom.
    \item $(1\times2)$: Single fragment with the 1s orbitals of both atoms.
    \item $(2\times2)$: Two fragments, each having the 1s and 2s orbitals of each atom.
\end{enumerate}

When comparing the one-shot calculations (right panel of Fig.~\ref{fig:H2ccpvdz}) with the self-consistent ones (left panel), it becomes apparent that both are qualitatively equivalent, and in fact are also in great quantitative agreement for most of the potential energy surface.
Infact, the main difference is the behaviour of the surfaces around the Coulson-Fisher point.
Since this is the point where an unrestricted mean-field treatment would break symmetries, it is perfectly reasonable that precisely in this regime the mean-field decoupling would have the most readjustment to do in response to the correlation in the Gutzwiller wave function.
In the bound and dissociation limits, the difference between the ground state energies in the one-shot and self-consistent variants is consistently below 0.1 mHa, well below chemical accuracy.
This is quite encouraging, as the self-consistent method can be significantly more expensive in computational ressources.
This suggests that for challenging systems where full self-consistency may be computationally prohibitive, a simple one-shot calculation has the potential of providing a qualitatively correct description of correlated phenomena such as bond breaking.

Comparing the potential energy surfaces for the different fragmentations, we recover a very similar picture to that obtained for the minimal basis H$_2$ calculations.
The single atom, single 1s orbital fragmentation $(1\times1)$ behaves essentially in a mean-field way towards the dissociation limit, although it does recover about half of the correlation energy in the bound region (see inset in Fig.~\ref{fig:H2ccpvdz}).
Correlating both 1s orbitals in separate fragments [$(2\times1)$] or in one and the same fragment [$(1\times2)$] significantly improves the dissociation limit, although a fix amount of correlation energy is missing.
This is likely due to charge fluctuations between the 1s and 2s orbitals in the atomic limit, since a fragmentation including both these orbitals in two atomic fragments [$(2\times2)$] exactly recover the right dissociation energy.
The $(2\times1)$ and $(1\times2)$ results significantly differ in the bound region, in which paradoxically the $(2\times1)$ fragmentation gives almost exact energies, while the $(1\times2)$ fragmentation is much more similar to the $(1\times1)$ one (and incidentally, the CASSCF results).
This is at first surprising, since the $(1\times2)$ fragmentation contains all interactions between the two atomic 1s orbitals explicitly, while the $(2\times1)$ one decouples them in mean-field.
In this case, it is likely a fortuitous error cancellation between the decoupled 1s-1s interactions, and the 1s-2s/1s-2p interactions lowering the energy of the $(2\times1)$ fragmentation.

All in all, the $(2\times2)$ fragmentation reproduces the potential energy surface with the best average accuracy across the inter-atomic distances.
While this is in good agreement with the intuition that only two orbitals should be correlated at any one point in the calculation, ideally one would find a single fragment embedding with two orbitals reproducing the full surface.
The CASSCF simulations with 2 electrons and 2 orbitals show that Gutzwiller is already describing the bound region accurately, since it perfectly agrees with the $(1\times2)$ fragmentation, while at the same time suggesting that the dissociation regime can be treated better: the CASSCF results can reproduce the exact dissociation energy with just two ``embedded'' orbitals.
The main difference between CASSCF and Gutzwiller is the orbital optimization inherent in the former.
The variational orbital optimization is effectively selecting which interaction terms are treated explicitly, and which decoupled in mean-field.
In this particular case, the CASSCF is likely a mixing of 1s and 2s orbitals in the dissociation limit, recovering all the atomic correlation energy.
While a similar orbital optimization can perhaps be devised for the Gutzwiller embedding, it does not seem to be necessary for a qualitative description at this stage.
And further, the $(2\times2)$ fragmentation seems to be doing better than CASSCF (2e, 2o), while only adding a polynomial overhead to the computational burden, i.e., solving two impurity models instead of one.

The mean-field + Gutzwiller Ansatz seems to be potentially capable of providing comparable results to established active space methods for ground state energies.
Now, we can turn our attention to one of particularities of the Gutzwiller framework: the possibility of inexpensively accessing spectral functions from the quasi-particle Hamiltonian.

\subsection{Hydrogen rings}

\begin{figure}
    \centering
    \includegraphics[width=0.5\textwidth]{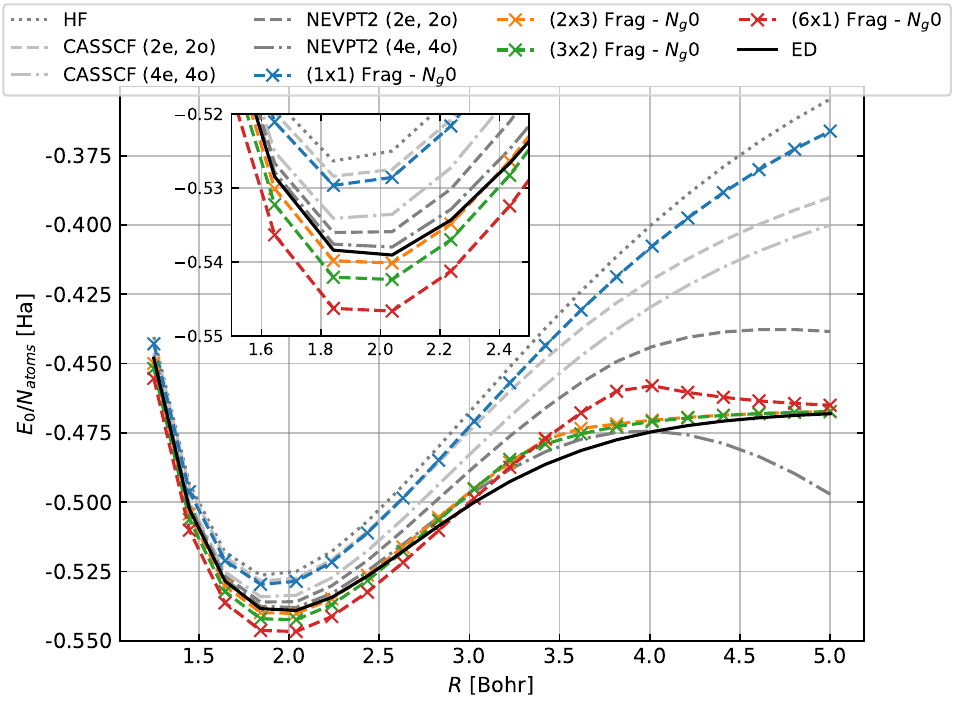}
    \caption{Potential energy surface for the dissociation of the H$_6$ ring in sto-3g basis, within the Gutzwiller approximation. Different fragmentations are tested, and compared to exact results and two different falvors of active space methods, see text for details.}
    \label{fig:H6pes}
\end{figure}

Our last example is the dissociation of a H$_6$ ring in sto-3g basis.
In the dissociation process, all H atoms are pulled apart at the same time.
This case presents a marked difference from the H$_2$ dissociation: here all orbitals are important to consider correlation, as in the dissociation limit each one will be exactly half-filled.
Consequently, we will first concentrate on the effect that different fragmentations have on the ground state energy.
Besides this, we will also carefully analyze the description of the spectral function of the molecule across the potential energy surface, and will unveil how both the number of ghosts $N_g$, but also the fragmentation, play a role to providing an accurate description.
As before, we choose the fragment orbitals from localized, atomic-orbital-like, orthonormalized rotations of the canonical orbitals obtained from a previous restricted HF calculation.
In the full molecule, there are 6 1s orbitals localized to each H atom.

\begin{figure*}
    \centering
    \includegraphics[width=0.7\textwidth]{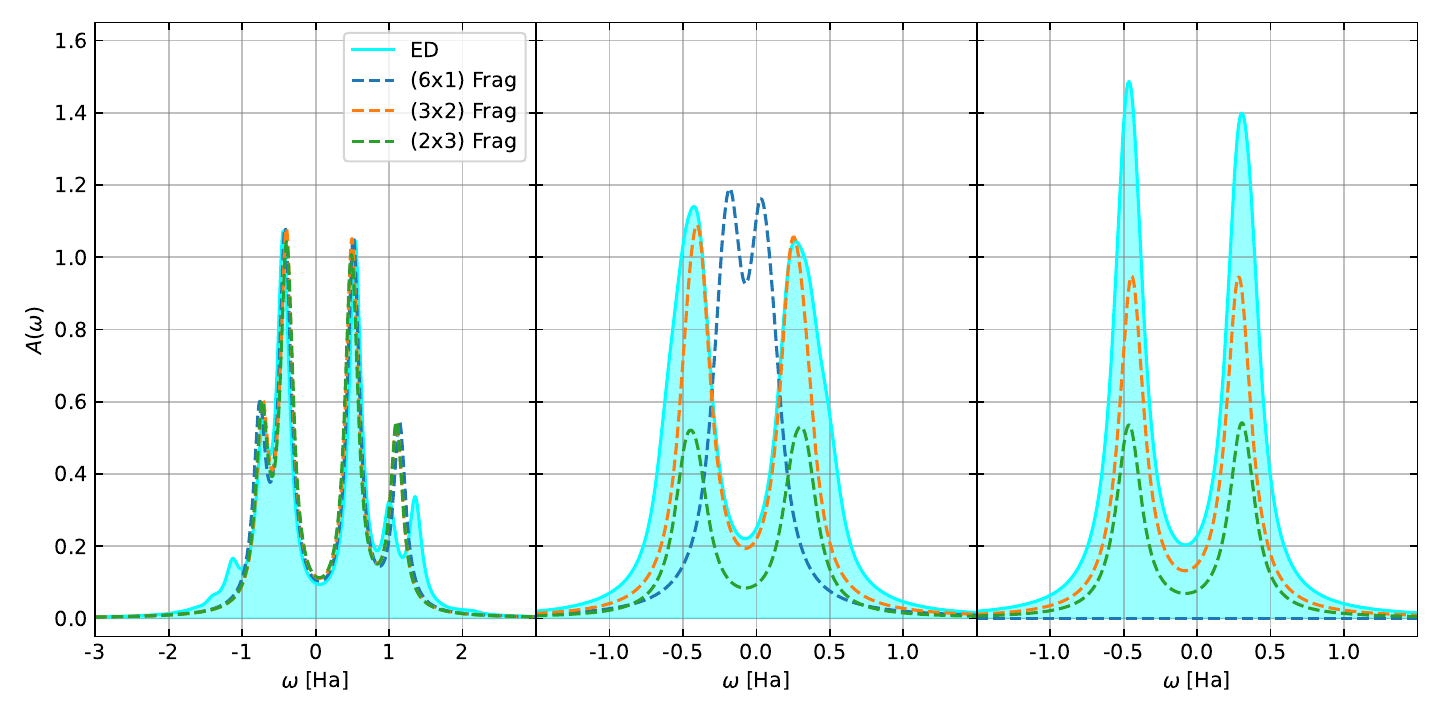}
    \caption{Spectral function of the H$_6$ ring in sto-3g basis for the equilibrium geometry (left), around $R = 3.8$ Bohr (Center), and in the dissociation limit (right). Computed with ED and Gutzwiller approximations for different fragmentations. See texts for detail.}
    \label{fig:H6_Spec_Frag}
\end{figure*}

Fig.~\ref{fig:H6pes} shows the potential energy surface for the H$_6$ dissociation for four different fragmentations within self-consistent mf+gGut:

\begin{itemize}
    \item $(1\times1)$: One fragment composed of the 1s orbital of only one of the six H atoms.
    \item $(6\times1)$: Six fragments, one for each 1s orbital.
    \item $(3\times2)$: Three fragments (dimers) composed from two 1s orbitals from neighboring H atoms.
    \item $(2\times3)$: Two fragments (trimers) composed from three 1s orbitals from neighboring H atoms.
\end{itemize}

These are in turn compared to exact results and two flavours of active space methods: CASSCF and CASSCF plus multi-reference perturbation theory (NEVPT2).
As expected, the $(1\times1)$ fragmentation essentially provides mean-field results.
All full-molecule fragmentations result in a reasonably faithful qualitative description of the full potential energy surface, including both bound and dissociation limits.
These compare favorably with CASSCF calculations with active spaces of (2e, 2o) and (4e, 4o), which share the main short-coming of the single fragment $(1\times1)$ embedding, that of completely missing the correlation in some atoms.
To address this, we also compare to NEVPT2 simulations.
Around the bound region, the largest trimer fragmentation is necessary to obtain comparable results to NEVPT2 (4e, 4o), but none of the NEVPT2 calculations can capture the dissociation behavior correctly.
The mf+gGut embedding can thus provide comparable, and on occasion better, results to sophisticated multi-reference correlation approaches.

Unsurprisingly, the larger the embedded fragments the better the overall description of the potential energy surface: the non-variational overestimation of the energy is reduced when moving from single atom to trimer embedding, and the bump of the Gutzwiller potential energy surface is also reduced in this direction.
Interestingly, however, for the regime around $R~\sim 3.5$ Bohr, the dimer fragmentation $(3\times2)$ is better than the trimer one $(2\times3)$.
This is likely related to the fact that the embedded dimers can build singlet states individually, while the trimers cannot, and have to build an overall singlet at the quasi-particle level.
In other words, around the bound region it is generally better to include more interactions explicitly into the model, whereas to capture the dissociation and electron localization it can be advantageous to have a fragment which can stabilize the right symmetries, in this case the spin state.
This notion, which is already suggested at the level of the energy, will become much more apparent when analyzing the spectral function $A(\omega)$.

\begin{figure*}
    \centering
    \includegraphics[width=0.7\textwidth]{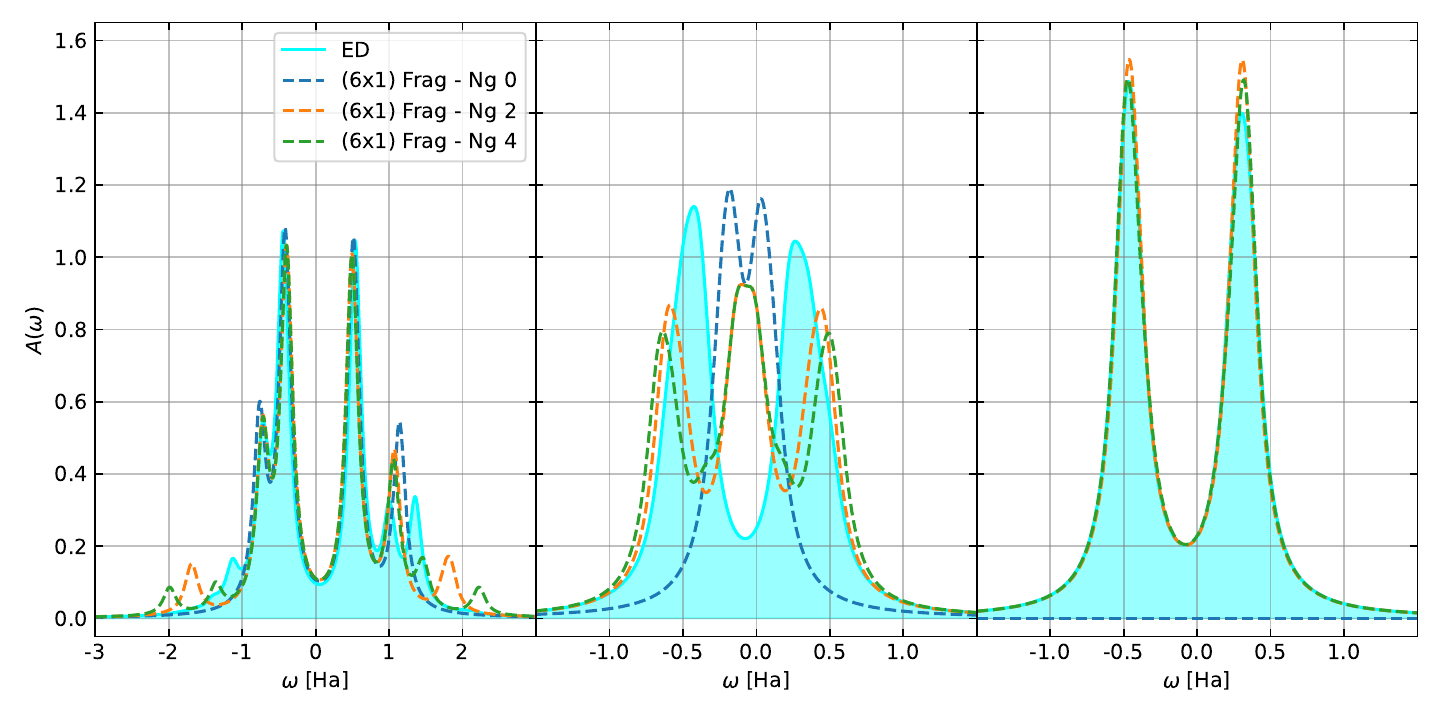}
    \caption{Spectral function of the H$_6$ ring in sto-3g basis for the equilibrium geometry (left), around $R = 3.8$ Bohr (Center), and in the dissociation limit (right). Computed with ED the ghost Gutzwiller approximations for the (6x1) fragmentation and different numbers of ghosts. See texts for detail.}
    \label{fig:H6_Spec_Ghost}
\end{figure*}

One of the main advantages of the Gutzwiller-based is that it gives access to the spectral function, which is related to (inverse) photo-emission spectra, in terms of the one-body quasi-particle Hamiltonian, see Eq.~\eqref{eq:gGut_GF}.
In this case, we can use the spectral function to analyze the nature of the Gutzwiller solution in the H$_6$ dissociation, and understand the model for bond breaking that it is introducing.
In Fig.~\ref{fig:H6_Spec_Frag} we present the spectral function for three different bond lengths, in the bound, intermediate and dissociated regions, for the different full-molecule fragmentations compared to the exact results.
In general, the picture follows quite closely what one would expect from the Brinkman-Rice scenario for the Mott transition: at the bound region, where electrons are mostly delocalized, all fragmentations provide reasonably accurate spectra functions. The satellites at $\sim-1.1$ Ha are not present, and the double peak at $\sim 1.2$ Ha is merged into one, but this can be partially remedied by introducing ghosts (see Fig.~\ref{fig:H6_Spec_Ghost}).
Towards the dissociation, when electrons localize, the atomic fragmentation ($6\times1$) closes the gap before ultimately losing all spectral weight after a first order transition, just like in the Gutzwiller picture of the Mott transition.
In contrast, the dimer and trimer embeddings keep a finite spectral weight throughout the full dissociation, and also qualitatively develop the correct spectral structure of two isolated peaks.
However, they too lose spectra weight, such that the spectral function is not normalized to one in this case either.
It seems thus that, regardless of the fragmentation, the Gutzwiller Ansatz loses some spectral weight when the electrons of the system localize.
This comes from the renormalization factors $R$ decreasing in magnitude to effectively decouple the embedded fragments from each other.
The right atomic limit is however buried below this harsh decoupling, as can be seen from the dimer and trimer results, so there is hope that adding ghosts will then recover the full spectral function in the dissociation limit, as was the case in the lattice models.

Fig.~\ref{fig:H6_Spec_Ghost} shows the spectral functions of H$_6$ for the same 3 configurations as Fig.~\ref{fig:H6_Spec_Frag}, but for the $(6x1)$ fragmentation with different numbers of ghosts $N_g$.
These are the number of ghosts in each fragment, such that the quasi-particle Hamiltonian in the case of $N_g = 4$ has 30 orbitals in total.
We see that the addition of ghosts greatly improves the spectral functions throughout.
In the bound region, the addition of ghosts improves the position o the shoulder at $\sim-0.8$ Ha, and adds satellite peaks to the spectrum in qualitatively the right regions.
Still, the satellites positions are not accurate, and while increasing the number of ghosts steadily improves them, it does so at the cost of adding spurious peaks.
After all, these correlated spectral functions come from the spectral function of the quasi-particle Hamiltonian, and hence will show in principle as many poles as orbitals in $H_{qp}$.

Nevertheless, the merit of the ghosts lies not on the fine tuning in the bound region, gGut is still a qualitative method.
It lies instead in the significant improvement of the spectra in the intermediate and dissociation regimes.
Adding ghosts essentially provides exact spectra in the dissociation limit, completely curing the spectral weight loss in the $N_g = 0$ simulation.
This is completely analogous to the case of the Mott transition of the Hubbard model in Fig.~\ref{fig:ghosts_example}.
While the intermediate regime also significantly improves, developing two distinct peaks much closer to the exact ones, there is also a pronounced peak that appears at zero frequency.
This is extremely evocative of a Kondo resonance in the Mott transition, and is indeed a spurious peak that is coming from the underlying impurity model description in gGut for the $(6\times1)$ fragmentation.
In this case, each atom is hybridized with a finite, non-interacting bath, exactly as happens in the DMFT picture of the paramagnetic Mott transition.
Towards dissociation, each of these atoms essentially decouple, and are populated by a single electron.
At the same time the total wave function of the impurity model corresponding to each atom has to be a singlet, given the current spin-independent (paramagnetic) formulation of the Ansatz, and this can only be done by hybridizing with the bath.
The electron localization and single-impurity-singlet condition in this embedding lead naturally to a Kondo-like resonance, which is the spurious peak that appears in the spectrum.
This clearly unveils the particular model with which the gGut Ansatz is attempting to describe the chemical bond breaking.
This spurious degree of freedom can finally be eliminated by choosing a fragmentation including more than a single atom per fragment.

\begin{figure}
    \centering
    \includegraphics[width=0.5\textwidth]{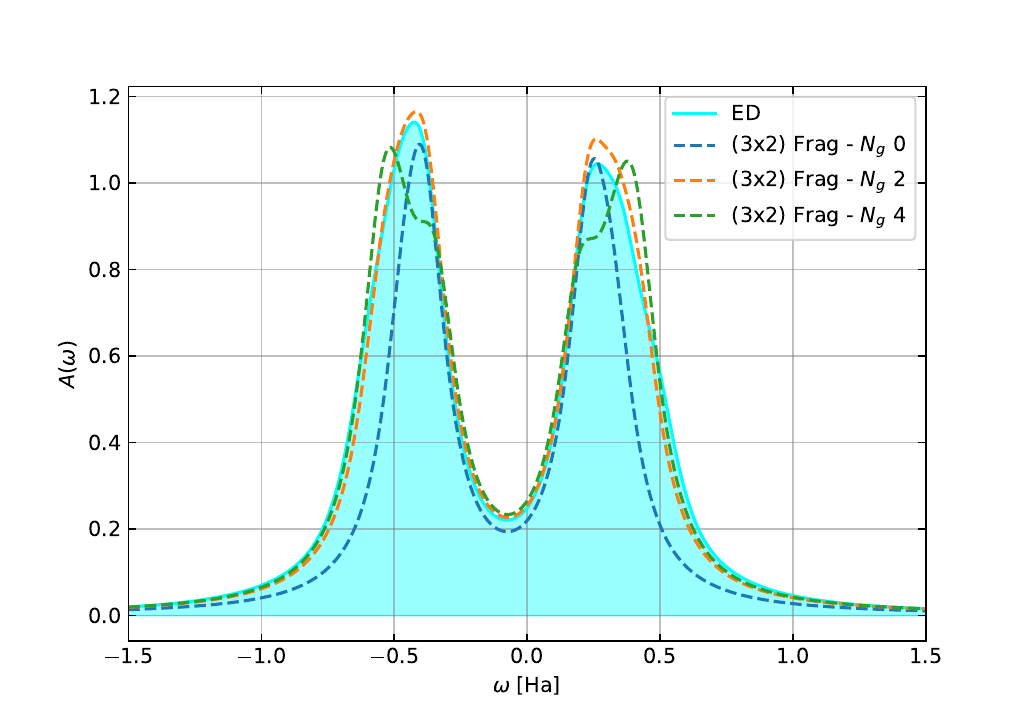}
    \caption{Spectral function of the H$_6$ ring in sto-3g basis around $R = 3.8$ Bohr. Computed with ED the ghost Gutzwiller approximations for the (3x2) fragmentation and different numbers of ghosts. See texts for detail.}
    \label{fig:H6_DimerGhosts}
\end{figure}

Indeed, in Fig.~\ref{fig:H6_DimerGhosts} we present the spectral function for the dimer fragmentation $(3\times2)$ for different numbers of ghosts.
In this case, the addition of ghosts does not introduce the spurious peak at zero frequency.
After all, the impurity models now include dimers, not single atoms, and these can form singlets in the half-filled limit without needing to involve the bath.
This analysis clarifies how the chemical bond breaking is modelled within the embedding perspective, and shows that the gGut Ansatz can capture the spectral functions of correlated molecular systems accurately.

Taking all different results into account, we have shown how the mf+gGut embedding framework can reliable capture the electronic correlation, both local and non-local, in correlated molecular systems with qualitative and sometimes even quantitative accuracy.
In terms of ground state energies, it can recover bond breaking potential energy surfaces faithfully, and in cases with great agreement with established complete active space methods.
Furthermore, upon convergence it provides access to accurate spectral functions in terms of the non-interacting quasi-particle Hamiltonian $H_{qp}$, despite the underlying self-consistency not invoking spectral information at any point.
This type of embedding shows hence promise for studying photo-emission spectra in correlated molecules which may be beyond the computational reach of more accurate multi-reference methods, as well as for proposing interpretable pictures for the origin of correlation in complex molecular systems, in terms of the underlying impurity models and quasi-particle Hamiltonian.

\section{Discussion - Beyond Mean-Field Decoupling of Non-Local Interactions}
The results of the previous sections show how the mf+gGut approximation can provide qualitatively correct descriptions of the non-local correlations in molecular systems, particularly within bond dissociation scenarios.
This includes potential energy surfaces, as well as spectral functions, which are related to measurable (inverse) photo-emission spectra.
Beyond this comparatively simple $\sigma$ bond dissociations, it will be interesting to pursue a thorough analysis of the quality of the ground state energy within the Ansatz in a broader range of systems, as well as investigating the convergence of correlated spectral features, such as shake-up satellites~\cite{Cederbaum1977,Cederbaum1980,Zhou2020,Mejuto2021,Marie2024,Loos2024}, with the number of ghosts.

Despite the encouraging successes in these initial calculations, some short-comings of the current approximation are clearly apparent.
Most notably, the loss of the variationality of the energy, which as discussed in the theory section is related to the multi-fragment version of the mean-field decoupling performed prior to the gGut embedding.
While a qualitatively correct picture of the electronic correlation seems still achievable, potential applications of this framework to systems presenting small energy gaps, such as the spin gaps in iron-sulphur clusters~\cite{Noodleman1988,Sharma2014,Mejuto2022}, will require a refinement of the methodology.

One potential option would be recovering the neglected electronic correlation in the mean-field decoupling, for instance by devising some perturbative correction.
Indeed, the gGut embedding is still, at its core, a variational wave function method, just not for the exact system Hamiltonian, but to an ``impurity-model-like'' version of it.
It should thus be possible to express the missing correlation due to the presence of full non-local two-body terms in the Hamiltonian as a perturbation on top of the gGut solution, and evaluate at least a first order correction to the energy by computing the expectation value of the corresponding perturbation operator.
This will require the introduction of renormalization tensors $\chi$ in Eq.~\eqref{eq:Xeq}, equivalent to the $R$ tensor which arises from evaluating the expectation value of non-local hoppings~\cite{Fabrizio2007} in Eq.~\eqref{eq:Req}.
Just like in the case of $R$, the value of $\chi$ can be determined from correlation functions in the impurity models alone~\cite{PasquaSoon}.
This would thus represent a first step into recovering the missing non-local correlations in the current implementation.

Ultimately, however, one should aim to rigorously include non-local interactions into the gGut Ansatz.
This means abandoning the mean-field decoupling before applying the embedding, and instead evaluating all relevant non-local expectation values which will arise in Eq.~\eqref{eq:GutAnsatz}.
This will result in three main differences with the current algorithm: (i) the presence of the renormalization tensors $\chi$ arising from the presence of non-local interactions, which will need to be included in the self-consistency, (ii) the emergence, through the tensors $\chi$, of impurity-bath interaction terms in the impurity Hamiltonians and (iii) the presence of non-local interaction terms in the quasi-particle Hamiltonian.
The effect of (i) on the convergence properties of the self-consistency have to be subject to their own study, but for the current discussion let us assume that a stable fix-point can still be found in reasonable time.
The second point, the appearance of impurity-bath interactions, does not need to be a major hindrance, as the spirit of any embedding approach relies in choosing a small enough impurity (fragment).
Hence, as long as obtaining an accurate representation of the 1-RDM of the impurity model is still computationally feasible, point (ii) is no issue of concern.
Indeed, for small enough fragments and numbers of ghosts, exact diagonalization (ED) methods are applicable regardless of the presence of bath-impurity interactions.
For larger impurity models than what ED can handle, a palette of different correlated approaches exists with a marked history of success in approximating 1-RDMs accurately at a moderate cost, e.g. selected configuration interaction methods~\cite{Zgid2011,Zgid2012,Lu2014,Go2015,Go2017,Mejuto2019,Werner2023,Williams-Young2023,Bellomia2024}, tensor network approaches~\cite{Schollwoeck2005,Schollwock2011,Nunez2014,Nunez2018,Wolf2015a,Bauernfeind2017,Lu2019,Paeckel2023}, or coupled-cluster based techniques~\cite{Zhu2019a,Zhu2019b,Shee2019,Zhu2020,Zhu2020b,Yeh2021,Zhu2021,Shee2023,damour2024}.

On the other hand, point (iii) poses a great challenge.
Indeed, the quasi-particle Hamiltonian has at least as many orbitals as the physical one, more in the case that ghosts are added.
Hence, making it a fully interacting Hamiltonian results in the ``quasi-particle'' problem being exactly as difficult to solve as the original one.
However, this is still a ``quasi-particle'' Hamiltonian, as all sources of local, i.e. in-fragment, electronic correlation have been relegated to the impurity models.
Therefore, approximating the effect of non-local correlations in $H_{qp}$ should be an acceptable trade-off, if the embedded fragments have been chosen judiciously.
The simplest such approximation would be a complete mean-field decoupling of the non-local interactions, which would lead us to an $H_{qp}$ very similar to the one in the mf+gGut method presented in this work.
Despite this apparent similarity, there are two main differences which make the approximation discussed in this section more accurate:
first, non-local interactions are only decoupled in mean-field at the quasi-particle level, but they still enter both the variational self-consistency, through the tensors $\chi$, as well as the impurity models, through the impurity-bath interactions.
In this sense, this mean-field decoupling should lose significantly less physical information than the strategy presented in the current work, which completely decouples all manners of non-local correlations in mean-field before establishing the embedding treatment.
Moreover, a full mean-field decoupling of all non-local interactions in $H_{qp}$ would still correspond to a variational approximation for the effect of non-local interactions, since these are all the interaction terms in this Hamiltonian.
Thus, this scheme will also remedy one of the main shortcomings of the present implementation: the loss of variationality in the energy (beyond that caused by the Gutzwiller approximation itself).
Furthermore, in cases where a mean-field decoupling is not accurate enough, one can again leverage the fact that only the 1-RDM and energy are needed for the self-consistency to use more complex approximations capturing more electronic correlation, such as perturbative schemes or coupled cluster methods.

\section{Conclusions}
Taking everything into consideration, the ghost Gutzwiller framework represents a promising alternative Ansatz to treat electronic correlation in molecular systems within quantum embedding, based on introducing auxiliary, non-interacting quasi-particle states to model correlation in terms of additional one-body fluctuations.
It can provide qualitatively accurate energies, as well as spectral functions in correlated molecules, with a transparent interpretation in terms of impurity models and moderate computational cost.
This work has shown an initial approximation to treat non-local correlations, which despite its simplicity correctly captures bond dissociation in toy models, the Hydrogen molecule and Hydrogen rings.
Moreover, we have discussed possible future directions to complement the formalism, and recover the main correlation contributions missing in the current approach.
While further work is necessary to assess the quality of the approximation in more challenging systems, such as transition-metal clusters, this initial study presents encouraging results suggesting that embedding using auxiliary quasi-particle states can become a useful tool in the study of strong correlation in molecules.
Particularly, the variational origin of the method allows the evaluation of forces, such that this embedding could find applications in systems where vibronic couplings are important.

\section*{Acknowledgements}
We thank Michele Fabrizio, Massimo Capone, Adriano Amaricci and Samuele Giuli for insightful discussions.

\end{document}